\documentclass[lettersize,journal]{IEEEtran}
\usepackage{amsmath,amsfonts,amsthm,amssymb,bbm}
\usepackage{algorithm}
\usepackage{algpseudocode}
\usepackage{array}
\usepackage[caption=false,font=tiny,labelfont=sf,textfont=sf]{subfig}
\usepackage{textcomp}
\usepackage{stfloats}
\usepackage{url}
\usepackage{verbatim}
\usepackage{graphicx}
\usepackage{bm}
\usepackage{subfig}
\usepackage{booktabs}
\usepackage{ulem}
\usepackage{color}

\def\BibTeX{{\rm B\kern-.05em{\sc i\kern-.025em b}\kern-.08em
    T\kern-.1667em\lower.7ex\hbox{E}\kern-.125emX}}
\usepackage{balance}

\begin{document}
\title{Dynamic Adaptive Resource Scheduling for Phased Array Radar: Enhancing Efficiency through Synthesis Priorities and Pulse Interleaving}
\author{Mingguang Han}

\maketitle

\begin{abstract}
To enhance the resource scheduling performance of phased array radar, we propose a dynamic adaptive resource scheduling algorithm based on synthesis priorities and pulse interleaving. This approach addresses the challenges of low efficiency, high loss ratios, and significant subjectivity in task assignment within phased array radar systems. We introduce a task synthesis priority design method that considers the working mode priority, deadlines, and time shift ratios. By implementing this method, we can increase the flexibility of task scheduling and improve the efficiency of radar processing tasks. Additionally, our proposed pulse interleaving method effectively utilizes the waiting periods between receiving and transmitting pulses to process other beams, thereby enhancing resource utilization. Simulation results demonstrate that the proposed scheduling algorithm significantly reduces time deviation ratios and scheduling failure rates while improving scheduling yield and time utilization ratios.
\end{abstract}


\section{Introduction}
\noindent Phased array radar exhibits rapid beam agility and effective waveform self-adaptation capabilities. It is capable of performing various tasks, such as multiple target tracking while simultaneously searching for new targets \cite{[1]Lu2005}. However, the multitasking capacity of phased array radar is constrained by limited time, energy, and computing resources. Therefore, implementing effective task priority planning methods and resource scheduling strategies is crucial for achieving optimal performance in multifunctional phased array radars. Several approaches have been developed for designing resource scheduling strategies for phased array radars \cite{[2]Cheng2008}, \cite{[3]Sathish2004}, \cite{[4]Zhang2009}, including fixed templates, online templates, partial templates, and adaptive algorithms. While template-based methods are simple to design, they often result in low task scheduling efficiency and resource utilization ratios. In contrast, adaptive resource scheduling algorithms leverage the flexibility of radar beam scanning, maximizing the use of time, energy, and other resources for efficient task scheduling. Currently, adaptive resource scheduling algorithms are widely adopted in the resource scheduling of phased array radars.

The adaptive resource scheduling algorithm is primarily divided into two modules: task priority design and scheduling algorithm design. The scheduling algorithm itself can be approached from two main perspectives \cite{[5]Huizing1996}. The first involves designing a scheduling algorithm based on the scheduling interval. Within the system's constraints of executable time and available resources, the optimal execution time for a task is selected according to specific criteria. The second perspective focuses on the time pointer, wherein the highest priority task that can be executed at a given moment is selected and added to the execution queue. Huizing \cite{[6]Mir2014} introduced the concept of a task time window, which allows the actual execution time of a scheduled task to shift within a defined time window around the expected execution time. This method significantly enhances the task scheduling success rate and time utilization of the radar system while mitigating the impact of time shifts on scheduling timeliness.

In terms of task priority design, resources in phased array radar (PAR) are allocated to tasks based on predefined rules, with typical strategies prioritizing the highest task working mode \cite{[7]Zhang1997}, \cite{[8]Kuo2005}, \cite{[9]Lu2006} and the earliest deadline \cite{[10]Wang2008}, \cite{[11]Sun2017}. However, these algorithms primarily rank tasks based on a single parameter, resulting in poor adaptability and low flexibility in scheduling. Recent studies [12-17] have proposed a two-factor approach that considers both the highest task working mode and the earliest deadline (HPEDF) to design task synthesis priorities. Furthermore, \cite{[18]Zhang2016c}, \cite{[19]Sourav2004} employed a broader set of characteristic parameters to establish a comprehensive task priority, mapping task working modes, deadlines, and idle times to a unified scale, thereby improving the scheduling performance of the algorithm. Additionally, \cite{[20]Yan2017} suggested incorporating a target threat factor in the design of task comprehensive priority, enhancing the efficacy of task scheduling.

References [21-23] employ a secondary planning algorithm focused on the scheduling interval to determine the optimal execution time of tasks. This approach aims to ensure that the actual execution time of tasks aligns closely with their expected execution time, thereby optimizing the execution timing for high-priority tasks. While these algorithms enhance the timeliness of radar system scheduling, they may reduce the time utilization and flexibility of task scheduling.

Most radar scheduling algorithms cited in other studies are designed from the perspective of time pointers. This approach involves selecting the highest priority task that can be executed at a specific moment and adding it to the scheduling execution queue. Algorithms designed with this methodology exhibit high flexibility, a high time utilization rate (TUR), and a high scheduling success rate (SSR). In traditional phased array radar systems, each mission is treated as a whole, preventing the scheduling of other tasks during the mission's designated time. However, advancements in pulse interleaving technology \cite{[24]Cheng2009} allow for the utilization of waiting periods between radar beam transmission and reception. Studies \cite{[13]Zhang2016}, [25-28] have integrated pulse interleaving technology into their scheduling algorithms, resulting in improved TUR and SSR.

Furthermore, references [29-32] have introduced intelligent algorithms, such as genetic algorithms, into the scheduling strategies to enhance scheduling efficiency. While genetic algorithms are shown to improve SSR and TUR, they also lead to increased time shift rates (TSR), which violates the expected timing principles and adversely affects the timeliness of the radar system. To enhance the flexibility of radar task scheduling, references [34-35] propose adjusting the scheduling intervals. Additionally, [36] introduced a resident scheduling algorithm that alters the radar's operating mode based on target distance.

Although these algorithms improve TUR and SSR to varying extents, they also result in increased TSR. This issue becomes particularly pronounced with the significant rise in tracking tasks [37-40], leading to higher TSR and consequently violating the expected timing principles of scheduling algorithms, thereby impacting the timeliness of radar operations. The threat levels of different targets and the priorities of various airspaces influence the operational modes of radar missions, making target threat levels and airspace priorities critical factors in mission priority design. Moreover, as the number of tracking tasks increases, the implementation of pulse interleaving technology can effectively enhance TUR and SSR. However, the challenge remains in efficiently interleaving radar tasks due to the constraints of energy resources within the radar system.

To address these challenges and enhance the overall scheduling performance and scheduling gain of radar systems, this paper presents a resource scheduling algorithm for phased array radar based on synthesis priorities and pulse interleaving. The main contributions of this study are as follows:
\begin{enumerate}{}{}
\item{Phased array radar mission structure, which consists of the transmit interval, the wait interval and the receive interval, and established a task scheduling optimization model under the constraints of time and energy resources. The objective function includes mission importance, urgency, and timeliness. }
\item{A comprehensive priority design method that comprehensively considers the target threat degree, airspace priority, task work mode, deadline and time shift rate is proposed. On the basis of improving scheduling success rate and time utilization of the algorithm, the time offset rate is effectively reduced, and the timeliness of the radar system is improved. The performance of the algorithm in many aspects can be effectively guaranteed.}
\item{A pulse interleaving algorithm that considers two interleaving modes under the constraints of time and energy resources, which follows the principle that the adjacent pulse receiving interval is close to each other, effectively reduces the computational complexity, and improves time utilization rate and scheduling success rate.}
\item{Combining the task synthesis priority design method and pulse interleaving technology to realize radar resource scheduling, the effectiveness and efficiency of this algorithm are verified by the simulation experiments of randomly generated tasks and the experimental results under different task scales.}
\end{enumerate}

\textbf{Outline}:
The organization of the remainder of this paper is as follows: Section II presents the radar task resident scheduling model, detailing the associated resource constraints and evaluation indicators. Section III describes the implementation methodology of the proposed adaptive resource scheduling algorithm, which is based on synthesis priorities and pulse interleaving. In Section IV, simulation experiments are conducted to compare and validate the proposed method against existing approaches. Finally, Section V concludes the paper by summarizing the key findings of this research.

\section{Modeling}
\subsection{MFPAR task model}
\noindent The overall scheduling structure of the Multifunctional Phased Array Radar (MFPAR) is illustrated in Figure 1. Initially, the radar system receives requests and generates radar tasks, such as search and tracking tasks. These tasks include parameters such as expected execution time, time windows, and beam dwell times. The scheduling algorithm organizes the requested tasks based on their priorities and the constraints of radar resources, categorizing them into execution queues, delay queues, and delete queues. Tasks in the execution queue are scheduled for execution according to a predefined execution list, while tasks in the delay queue are resubmitted to the scheduler for reconsideration during the next update cycle. Tasks in the delete queue are discarded.
		\begin{figure}[!t]
		\centering
		\includegraphics[width=3in]{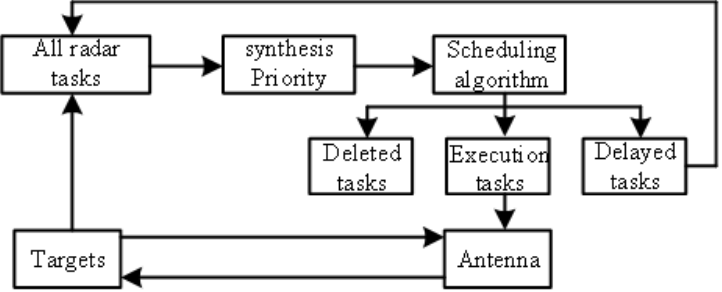}
		\caption{Block diagram of task scheduling for MFPAR.}
		\label{fig:1}
		\end{figure}

Most adaptive scheduling algorithms focus primarily on the time resources of the radar system, often overlooking constraints related to energy and computational resources. While it can be argued that the bottleneck posed by computational resource limitations is gradually diminishing due to advancements in computer performance, the energy resource constraints of the radar system remain critical, particularly as radar cannot remain in a transmitting state for extended periods. The implementation of a pulse interleaving strategy allows the waiting periods between radar pulses to be utilized for transmitting other radar pulses.

Tracking tasks can be classified according to the threat levels associated with operational conditions, which may include high-precision tracking, precision tracking, ordinary tracking, or simple monitoring. Figure \ref{fig:2} illustrates the task model of pulsed phased array radar, which consists of three components: the launch period, waiting period, and reception period. The i-th phased array radar task can be described as follows:
		\begin{figure}[!t]
		\centering
		\includegraphics[width=3in]{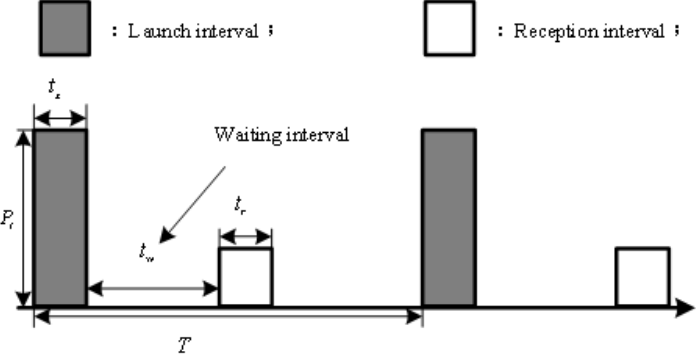}
		\caption{The working model of Phased array radar pulse.}
		\label{fig:2}
		\end{figure}

\begin{equation}\label{eq:1}
T_i = \{P_{ri}, t_{ei}, \eta_{i}, t_{si}, \Delta t_{i}, l_{i}, t_{xi}, t_{wi}, t_{ri}, t_{ci}, t_{di}, \Delta T_{i}, P_{ti}\}
\end{equation}

In ~\eqref{eq:1}, \(P_{ri}\) represents the task type, \(t_{ei}, \eta_{i}, t_{si}\) stand for the task request time, task attribute, and actual execution time respectively. \(\Delta t_{i}\) represents the beam dwell time of the task. \(l_{i}\) is the time window length, \(t_{xi}\) for the transmit interval, \(t_{wi}\) for the wait interval, which depends on the target distance, and \(t_{ri}\) for the receive interval. \(\Delta t_{i}\) satisfies:
\[
\Delta t_{i} = t_{xi} + t_{wi} + t_{ri}
\]

where \(t_{ci}\) is the pulse launch cooling time, and \(t_{di}\) is the deadline. Each task should be executed before its deadline, \(t_{di}\) satisfies:
\[
t_{di} = t_{ei} + l_{i}
\]

\(\Delta T_{i}\) is the sample interval between two adjacent tasks of the same type. It satisfies:
\[
t_{ei} = t_{s(i-1)} + \Delta T_{i}
\]

where \(t_{s(i-1)}\) denotes the execution time of the former task. \(P_{ti}\) is the task power consumption. In reality, the transmit interval \(t_{xi}\) consumes the majority of the power of the whole task.

As the task scheduling of each working mode needs to be realized through the radar pulses, it will consume certain system energy and time resources. Considering that the radar cannot always be in the transmitting state, energy resource constraints should be applied to the phased array radar when the pulse interlace strategy is used.

In order to maximize resource utilization and scheduling efficiency, various resource constraints must be considered when performing resource scheduling.

\subsection{Resource constraints}
1) Energy Resource Constraint

Radar systems are well-known to consume a significant amount of energy during pulse transmission. When pulse streams are interlaced, the operational time of the radar transmitter is extended, leading to increased energy consumption. If the operating temperature of the radar exceeds the physical limits of the system, there is a risk of damaging the transmitter. During the tracking process, to maintain adequate tracking quality, the width of the transmitted pulse must be increased, which further escalates energy consumption. It has been observed that energy constraints are influenced by the selected modes of operation, which impacts scheduling analysis in the following ways:
\begin{equation}\label{eq:2}
P_{\tau}(t) \leq P_{\tau_{\text{max}}}
\end{equation}

In \eqref{eq:2}, \(P_{\tau_{\text{max}}}\) is the upper limit of radar instantaneous power consumption. \(P_{\tau}(t)\) is the power consumed by the radar transmitting a pulse at time \(t\).
\begin{equation}\label{eq:3}
P_{\tau}(t) = \frac{1}{\tau} \int_{-\infty}^{t} p(x)e^{(x-t)/\tau} \, dx
\end{equation}

The cooling time constant is \(\tau\), which characterizes the heat dissipation performance of the radar. The energy consumed during a radar receiving period is negligible, and the time difference between the end of a pulse emission period and the start time of the next pulse transmission period is \(t_{ci}\). It should be ensured that the energy at the end of the next scheduled mission launch period does not exceed the instantaneous maximum power.
\begin{equation}\label{eq:4}
t_{ci} = -\tau \ln \left( \frac{P_{\tau_{\text{max}}} - A_{i}}{P_{\tau_{\text{max}}} e^{-t_{i}/\tau}} \right)
\end{equation}

2) Time Resources Constraint

Because radar scheduling tasks are mainly performed at regular scheduling intervals, once the scheduling interval is determined, the radar task scheduling becomes less flexible. The launch and reception periods of radar missions cannot be preempted, and all missions must be completed before their deadlines. In a traditional scheduling algorithm, a single beam dwell period cannot be preempted, and the number of tasks, \(N_{1}\), successfully executed within one scheduling interval must satisfy:
\begin{equation}\label{eq:5}
\sum_{k=1}^{N_{1}} \Delta t_{k} \leq t_{sI}
\end{equation}

In \eqref{eq:5}, \(\Delta t_{k}\) is the length of the beam dwell period, and \(t_{sI}\) is the duration of the scheduling interval. In order to improve the utilization of radar time resources, the pulse interleaving strategy is employed to interlace other tasks during the waiting period between each radar task reception and transmission. The pulse transmission period and reception period of a task cannot be preempted by other tasks; otherwise, the task execution will fail. Successful execution of \(N_{i}\) tasks imposes the following constraints:
\begin{equation}\label{eq:6}
\bigcap_{i=1}^{N_{i}} [t_{s_{i}}, t_{s_{i}} + t_{x_{i}}] \cup [t_{s_{i}} + t_{x_{i}}, t_{s_{i}} + t_{x_{i}} + t_{w_{i}}] = \emptyset
\end{equation}
\begin{equation}\label{eq:7}
rt_{i2} + l_{2} \geq t_{end}, \; i_{2} = 1, \ldots, N_{2} 
\end{equation}
\begin{equation}\label{eq:8}
rt_{i3} + l_{3} < t_{end}, \; i_{3} = 1, \ldots, N_{3}
\end{equation}

\section{Task synthesis priority design}
\noindent In this paper, we employ a discrete Dynamic Bayesian Network algorithm to evaluate multi-target threats. Based on the assessed threat levels, tracking mode tasks are categorized into precise tracking and normal tracking. The importance of the radar mission and its associated deadline are utilized for scheduling purposes. Using a time-pointer algorithm, each task assignment necessitates a reprioritization process, allowing for dynamic evaluation of both the importance and urgency of radar tasks. This approach effectively reflects the changing significance of radar tasks over time.

Building upon the urgency factor, the classic Highest Priority Earliest Deadline First (HPEDF) synthesis priority design is adjusted by incorporating a time offset ratio. A more comprehensive radar task interleaving model is constructed, and we propose a synthesis task priority online pulse interleaving scheduling algorithm that incorporates the time-shift ratio. Finally, a series of simulation experiments are conducted to demonstrate the efficiency of the proposed algorithm.

\subsection{Task dynamic priority design}

In the task priority design process, this paper proposes an approach that first considers the task working mode, which reflects the importance of the task, and the task deadline, which indicates the urgency of the task. A priority table is designed to evaluate the importance and urgency of each task simultaneously, assigning them equivalent weights. Building upon this framework, a time deviation ratio is incorporated to enhance the efficiency of task scheduling, ensuring that critical tasks are executed as promptly as possible within their expected execution times. Ultimately, a synthesis priority design method is developed that accounts for task working modes, deadlines, and time shift ratios.
		\begin{figure}[!t]
		\centering
		\includegraphics[width=3in]{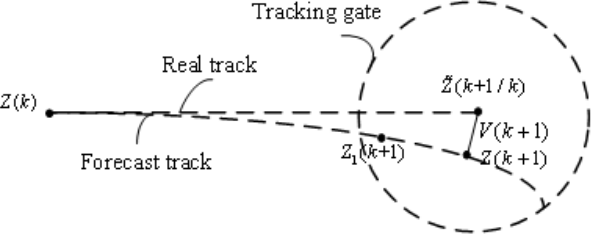}
		\caption{Schematic diagram of time shift and tracking accuracy.}
		\label{fig:3}
		\end{figure}

As shown in \ref{fig:3}, the \(k\)-th tracking target position is \(Z(k)\), the predicted value of the \(k+1\)-th tracking target position is \(\hat{Z}(k+1/k)\), the target measurement position is \(Z(k+1)\), and the new information is \(v(k+1)\). When there is a time offset, the target measurement position also changes; the new measurement position of the target will be recorded as \(Z_{1}(k+1)\), and the new interest is \(v_{1}(k+1)\). Consequently, the target tracking error will increase, and the new information will grow to a certain extent. Meanwhile, the target will fall into the tracking wave door with a greater probability, which may lead to tracking failure. Since the target tracking is filtered by a tracking algorithm such as an extended Kalman filter, the effectiveness of the target tracking is a quadratic function of the time offset \cite{[13]Zhang2016}. 

Therefore, in the comprehensive task priority design, the time shift should be synthesized into the task priority in the form of a quadratic function, maximizing the benefits of task scheduling and improving the effectiveness of task scheduling.

It is necessary to clarify the basic principles of task scheduling: (1) the higher the priority of the work mode, the higher the priority of the task; (2) the earlier the deadline, the higher the priority of the task; (3) the lower the time shift ratio, the higher the priority of the task.

\textbf{Step 1:} The overall priority of the task to be scheduled is 
\begin{equation}\label{eq:9}
p_{i} = f_{i} \cdot g_{i}
\end{equation}

Suppose that the current time scheduler has \(Q\) requests, which are recorded as \(q = \{q_{1}, q_{2}, \ldots, q_{Q}\}\), satisfying: (1) the arrival time of all requests is not greater than the current time; (2) the deadline of all requests minus the residence time is not less than the current time. 

First, the priority in accordance with the mode of operation of the size of the cut-off request \(Q\) tasks are arranged twice. Each task number is obtained in the arrangement, referred to as priority number \(N_{p_{i}}\), and the deadline \(N_{d_{i}}\), where \(1 \leq N_{p_{i}}, N_{d_{i}} \leq Q\). Then,
\begin{equation}\label{eq:10}
f_{i} = \frac{[(Q + 2 - \eta) \cdot N_{d_{i}} + \eta \cdot N_{p_{i}}]}{(Q + 1)}
\end{equation}
where \(\eta\) is a weighted value in the range of \([1, Q]\).

\textbf{Step 2:} The accuracy of tracking targets has a quadratic function relationship with the time shift. In order to make each scheduling task as close to its expected execution time as possible, the closer the time pointer is to the expected execution time, the higher the task priority is.

If the current request scheduling task operates in one of three modes: validation, precision tracking, and normal tracking, then
\begin{equation}\label{eq:11}
g_{i} = \begin{cases} 
\left( 1 - a \cdot \frac{|t_{ei} - tp|}{l_{i}} \right)^{2}, & tp \geq t_{ei} \\ 
\left( 1 - \frac{|t_{ei} - tp|}{l_{i}} \right)^{2}, & tp < t_{ei} 
\end{cases}
\end{equation}

If the current working mode of the request scheduling task is airspace search or horizontal search, then
\begin{equation}\label{eq:12}
g_{i} = \begin{cases} 
b \cdot \left( 1 - a \cdot \frac{|t_{ei} - tp|}{l_{i}} \right)^{2}, & tp \geq t_{ei} \\ 
b \cdot \left( 1 - \frac{|t_{ei} - tp|}{l_{i}} \right)^{2}, & tp < t_{ei} 
\end{cases}
\end{equation}

In Eq.\eqref{eq:11} and Eq.\eqref{eq:12}, the time pointer \(tp\) points to the current time \(\alpha \in (0, 1), b \in (0, 1)\). As the time window of validation and task tracking is smaller than that of task searching, the priority of validation and task tracking is higher than that of task searching under the same time shift ratio.

\subsection{Evaluation index}
This section proposes a method to evaluate the performance of the proposed scheduling algorithm:

\textbf{Successful Scheduling Ratio}

The Successful Scheduling Ratio (SSR) is defined as the ratio between the number of successfully scheduled tasks and the total number of request tasks. It can be expressed as:
\begin{equation}\label{eq:13}
S_{SR} = \frac{N_{1}}{M}
\end{equation}

In \eqref{eq:13}, \(N_{1}\) is the number of successfully scheduled tasks, and \(M\) is the total number of request tasks. The more tasks are successfully scheduled, the better the performance of the algorithm is.

\textbf{Time Utilization Ratio}

The Time Utilization Ratio (TUR) is the ratio of the launch duration and reception duration of all successfully scheduled tasks to the total time resource, which can be expressed as:
\begin{equation}\label{eq:14}
T_{UR} = \frac{1}{T_{0}} \sum_{i=1}^{N} (t_{xi} + t_{ri})
\end{equation}

where \(T_{0}\) is the total available time resource, and \(N\) is the number of successfully scheduled tasks. The higher the time utilization ratio in a phased array radar, the better the scheduling algorithm.

\textbf{Average Time Shift Ratio}

The Average Time Shift Ratio (ATSR) is the mean of the difference between the actual execution time of the successful scheduling tasks and the ideal execution time of them. This metric evaluates the effectiveness of task scheduling.
\begin{equation}\label{eq:15}
ATSR = \frac{1}{N} \sum_{i=1}^{N} \left| \frac{t_{si} - t_{ei}}{l_{i}} \right| 
\end{equation}

where \(N\) is the number of scheduled successful tasks, \(t_{si}\) is the actual execution time, \(t_{ei}\) is the task execution time, and \(l_{i}\) is the time window length of the task.

\textbf{Scheduling Yield Ratio}

The Scheduling Yield Ratio (SYR) is the ratio of the total yield of all tasks successfully scheduled to the ideal yield of all tasks. Since most classical Kalman filtering algorithms are nonlinear filtering algorithms, the target tracking accuracy is nonlinear with the time shift, and consequently, the effectiveness of the radar task scheduling is expressed as:
\begin{equation}\label{eq:16}
\left( 1 - k \cdot \left| \frac{t_{si} - t_{ei}}{l_{i}} \right| \right)^{2}
\end{equation}

Then the Scheduling Yield Ratio can be expressed as:
\begin{equation}\label{eq:17}
S_{YR} = \sum_{i=1}^{N} p_{i} \cdot \left( 1 - k \cdot \left| \frac{t_{si} - t_{ei}}{l_{i}} \right| \right)^{2} \bigg/ \sum_{i=1}^{N + N} p_{i}
\end{equation}

The importance of the various task types is expressed by \(k\), which could be \(1/2\).

\section{Resource scheduling strategy of phased array radar based on synthesis priority and pulse interleaving algorithm}
In the proposed task scheduling framework, the pulse interleaving strategy primarily utilizes time pointers for task allocation in a sequential manner. By allowing the waiting periods between tracking-type task pulses to transmit or receive other task pulses, the utilization of time resources can be significantly enhanced. To increase the task scheduling yield, improve scheduling effectiveness and time utilization, fully leverage the system's energy resources, and reduce computational demands, this paper builds upon the traditional HPEDF synthesis task priority planning algorithm.

Based on this foundation, we propose a dynamic task priority design algorithm that comprehensively considers task working modes, deadlines, and time offset ratios. This approach aims to ensure that both high and low-priority tasks are executed as closely as possible to their expected execution times while optimizing the utilization of time resources.

Additionally, this study employs a discrete Dynamic Bayesian Network algorithm to assess multi-target threats. Based on the evaluated threat levels, tracking mode tasks are categorized into precise tracking and normal tracking.

The scheduling of radar missions is guided by the importance of the mission and its associated deadline. Utilizing a time-pointer algorithm, each task assignment necessitates a reprioritization process. This approach allows for the dynamic evaluation of both the importance and urgency of radar tasks, effectively reflecting their significance over time. Based on the urgency factor, the classic HPEDF synthesis priority design is adjusted by incorporating a time offset ratio.

In this paper, we construct a more effective radar task interleaving model and propose a synthesis task priority online pulse interleaving scheduling algorithm that incorporates the time-shift ratio. Finally, a series of simulation experiments are conducted to validate the efficiency of the proposed algorithm.

\subsection{Scheduling strategy based on improved dynamic pulse interleaving}

To maximize the utilization of time resources and enhance the success ratio of radar task scheduling, it is essential to implement a pulse interleaving strategy. Building upon the dynamic synthesis priority model proposed in this paper, we develop a dynamic pulse interleaving algorithm designed to mitigate the complexity introduced by the integration of pulse interleaving.
		\begin{figure}[!t]
		\centering
		\includegraphics[width=3in]{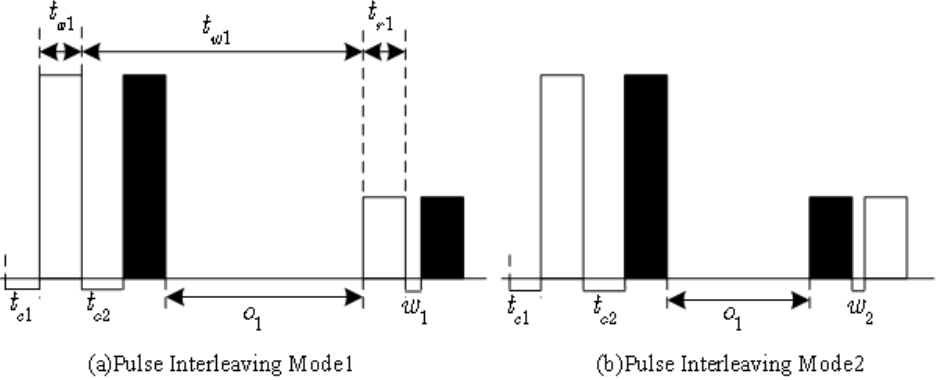}
		\caption{Two pulse interleaving methods for radar missions.}
		\label{fig:4}
		\end{figure}

Figure \ref{fig:4} shows two pulse interleaving modes for tracking tasks. Figure \ref{fig:4} (a) illustrates an external interleaving type pulse interleaving mode, while Figure \ref{fig:4} (b) depicts an internal interleaving type pulse interleaving mode. The blank rectangle represents the radar task 1, and the black rectangle represents the radar task 2. \(t_{1}, t_{2}\) are the cooling times of the radar transmitter. The \(o_{1}\) indicates the internal idle time of the pulse after pulse interleaving. 

The design principle of pulse interleaving between task 1 and task 2 is as follows: for impulse interleaving mode 1, minimize \(w_{1}\) as much as possible; for pulse interleaving mode 2, minimize \(w_{2}\) as much as possible, such that the reception of the adjacent radar pulses reception periods \(t_{r1}\) and \(t_{r2}\) is as close as possible.

The time must satisfy the following equations:
\begin{equation}\label{eq:18}
\begin{aligned}
(a) & \quad
\begin{cases}
o_{1} = t_{w1} \\
o_{1} - t_{c2} \geq t_{x2} \\
t_{r1} \leq t_{w2} \\
t_{c2} + t_{x2} + t_{w2} \geq t_{w1} + t_{r1}
\end{cases} \\
(b) & \quad
\begin{cases}
o_{1} = t_{w1} \\
o_{1} - t_{c2} \geq t_{x2} + t_{w2} + t_{r2}
\end{cases} 
\end{aligned}
\end{equation}

Assuming that there are \(N\) tasks waiting to be scheduled at time period \([t_{s(1)}, t_{end}]\), the specific steps of pulse interleaving are as follows:

\textbf{Step 1:} Initialize the time pointer \(tp = t_{0}\), the number of tasks that are scheduled \(q = 0\), radar pulse transmit power \(A_{1}\), previous radar pulse cooldown time \(t_{c} = 0\), radar pulse launch period end time power \(P_{out} = 0\), and pulse internal idle available time after pulse interleaving is \(o_{1} = 0\), and the pulse external idle available time is \(o_{2} = [t_{s(1)}, t_{end}]\).

\textbf{Step 2:} Select all the tasks that satisfy $tp \subset [t_{ei} - l_{i}, t_{ei} + l_{i} - \Delta t_{i}]$ and dynamically apply the synthesis priority design method according to \eqref{eq:9} to obtain a dynamic priority for each task and select the task with the highest priority.

\textbf{Step 3:} Determine whether the task to be scheduled can be pulse-interlaced with the task from the previous moment.

Case (1): If pulse interleaving is possible, the parameters after updating the interlaced task are as follows:
\begin{equation}\label{eq:19}
\begin{cases}
t_{ci} = -\tau \ln \left( \frac{P_{max} - A_{i} \left( 1 - e^{-t_{xi}/\tau} \right)}{P_{out}(i) e^{-t_{xi}/\tau}} \right) \\[2ex]
t_{s(i)} = t_{s(i-1)} + t_{x(i-1)} + t_{ci} \\[2ex]
tp = t_{s(i)} + t_{xi} \\[2ex]
P_{out}(i) = P_{out(i-1)} e^{-\left( \frac{t_{ci}+t_{xi}}{\tau} \right)} + A_{i} \left( 1 - e^{-t_{xi}/\tau} \right) \\[2ex]
o_{1} = o_{1} - t_{ci} - t_{xi} \\[2ex]
o_{2} = [t_{s(i)} + \Delta t_{j}, t_{end}] \\[2ex]
q = q + 1
\end{cases}
\end{equation}
\begin{equation}\label{eq:20}
\begin{cases}
t_{ci} = -\tau \ln \left( \frac{P_{max} - A_{i} \left( 1 - e^{-t_{xi}/\tau} \right)}{P_{out}(i) e^{-t_{xi}/\tau}} \right) \\[2ex]
t_{s(i)} = t_{s(i-1)} + o_{1} + t_{x(i-1)} - \Delta t_{i} \\[2ex]
tp = t_{s(i)} + t_{xi} \\[2ex]
P_{out}(i) = P_{out(i-1)} e^{-\left( \frac{t_{ci}+t_{xi}}{\tau} \right)} + A_{i} \left( 1 - e^{-t_{xi}/\tau} \right) \\[2ex]
o_{1} = t_{wi} \\[2ex]
o_{2} = [t_{s(i-1)} + \Delta t_{j}, t_{end}] \\[2ex]
q = q + 1
\end{cases} 
\end{equation}

Case (2): If the task to be scheduled is a tracking task and cannot perform pulse interleaving, update the parameters as follows: 
\begin{equation}\label{eq:21}
\begin{cases}
t_{ci} = -\tau \ln \left( \frac{P_{max} - A_{i} \left( 1 - e^{-t_{xi}/\tau} \right)}{P_{out}(i) e^{-t_{xi}/\tau}} \right) \\[2ex]
t_{s(i)} = tp \\[2ex]
tp = t_{s(i)} + t_{x(i)} \\[2ex]
P_{out}(i) = P_{out(i-1)} \cdot e^{-\left( \frac{t_{ei}+t_{xi}}{\tau} \right)} + A_{i} \left( 1 - e^{-t_{xi}/\tau} \right) \\[2ex]
o_{1} = t_{wi} \\[2ex]
o_{2} = [t_{s(i)} + \Delta t_{j}, t_{end}] \\[2ex]
q = q + 1
\end{cases}
\end{equation}

Case (3): If the task to be scheduled is different and it cannot interlace, the parameters are updated as follows:
\begin{equation}\label{eq:22}
\begin{cases}
t_{ci} = -\tau \ln \left( \frac{P_{max} - A_{i} \left( 1 - e^{-t_{xi}/\tau} \right)}{P_{out}(i) e^{-t_{xi}/\tau}} \right) \\[2ex]
t_{s(i)} = tp \\[2ex]
tp = t_{s(i)} + \Delta t_{i} \\[2ex]
P_{out}(i) = P_{out(i-1)} \cdot e^{-\left( \frac{t_{ei}+t_{xi}}{\tau} \right)} + A_{i} \left( 1 - e^{-t_{xi}/\tau} \right) \\[2ex]
o_{1} = 0 \\[2ex]
o_{2} = [t_{s(i)} + \Delta t_{j}, t_{end}] \\[2ex]
q = q + 1
\end{cases}
\end{equation}

Case (4): If there is no task request execution at the current time, update the time pointer \(tp = tp + dt\), where \(dt\) is the minimum moving step size.

\textbf{Step 4:} When \(tp > t_{end}\) or \(q > N\), go to Step 6; otherwise, return to Step 2.

\textbf{Step 5:} Delete the task that is not scheduled for execution, obtain the schedule execution linked list, delete the linked list, and end the task scheduling analysis in the scheduling interval.

\begin{algorithm}[!t]
\caption{Task Scheduling Process}
\label{alg:Task_Scheduling_Process}
\begin{algorithmic}[1]
\State \textbf{Initialize:} parameters
\If {shutdown or end simulation} 
    \State End scheduling analysis
\Else
    \State Take the events in the current scheduling interval and place them into queue \(T\). Set a total of \(N\) tasks, \(i = 0\).
\EndIf

\While {not shutdown or end simulation}
    \State Select the task where the time pointer \(tp\) falls in the interval \([t_{ei} - l_{i}, t_{ei} + l_{i} + \Delta t_{i}]\)
    \State Perform task comprehensive priority evaluation.

    \State Select the task that meets the highest overall priority at the current time.

    \If {the task to be scheduled can be pulse interleaved}
        \State Schedule the execution of the task at time \(tp\)
        \State Update the parameters according to equations (19) and (20).
    \Else
        \If {the task to be scheduled is a tracking task}
            \State Schedule the execution of the task at time \(tp\)
            \State Update the parameters according to equation (21).
        \Else
            \State Schedule the execution of the task at time \(tp\)
            \State Update the parameters according to equation (22).
        \EndIf
    \EndIf

    \If {$i \geq N$ or $t_{p} \leq t_{end}$}
        \State break
    \EndIf
\EndWhile
\end{algorithmic}
\end{algorithm}

Fig. 4 shows the algorithm scheduling process. When performing task scheduling, it satisfies the time constraints in  \cite{eq:18}, \(P_{out} \leq P_{max}\) energy constraint, and the dynamic scheduling update of allocation scheduling tasks on the timeline in the form of time pointers, which simplifies the complexity of scheduling analysis.
		\begin{figure*}[!t]
		\centering
		\includegraphics[width=6in]{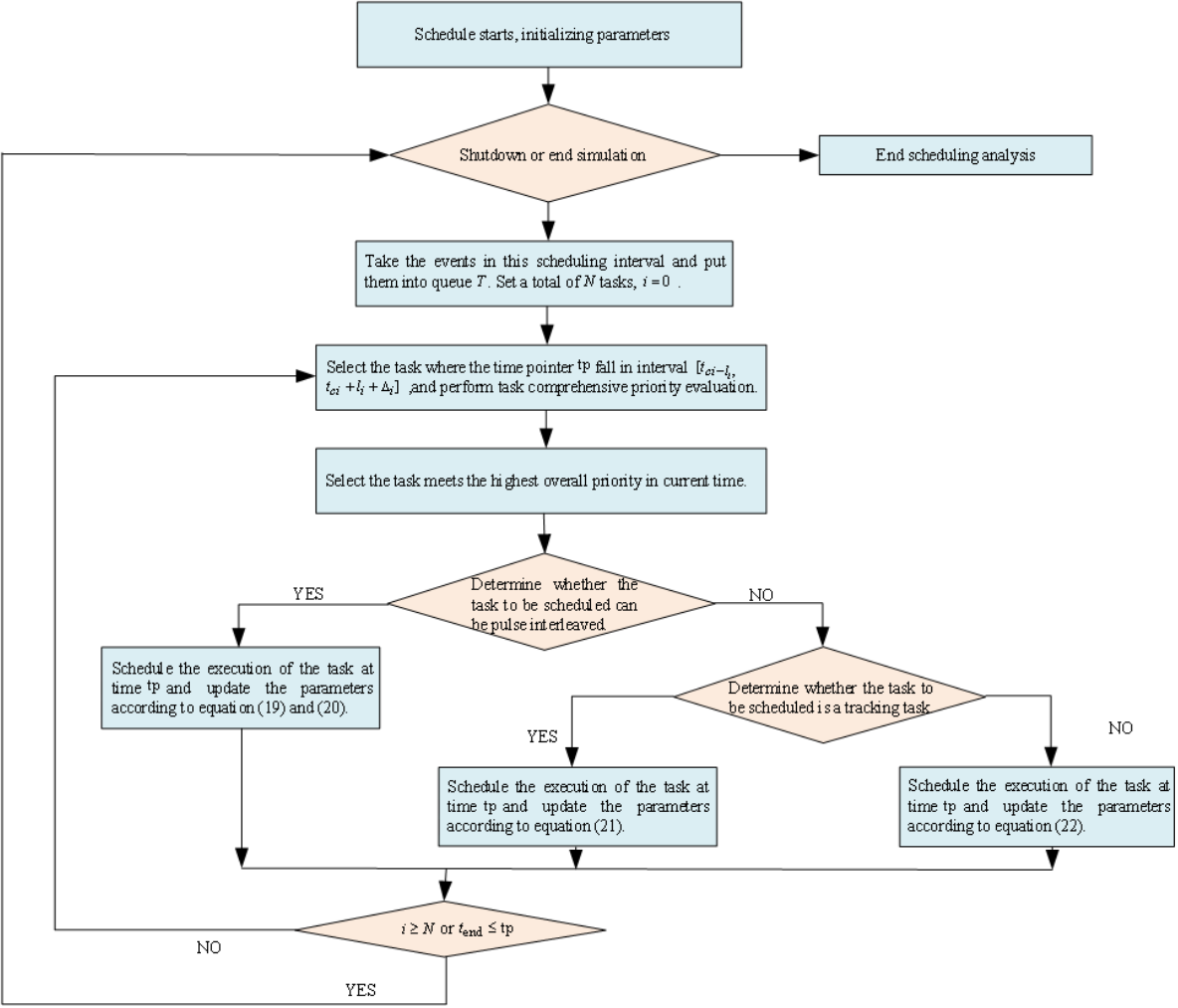}
		\caption{Algorithm scheduling process.}
		\label{fig:5}
		\end{figure*}

\section{Simulation experiment}
\subsection{Simulation Setup.}

The simulation scenario includes five types of tasks to be scheduled: verification, precision tracking, general tracking, horizon search, and airspace search. Task tracking is categorized into precision tracking and general tracking based on the level of target threat. The verification radar task consists of two components: one for generating verification tasks for target detection, and the other for generating false alarm tasks at a specific data ratio. The number of general target tracking tasks is fixed at 40, while the number of precision tracking tasks gradually increases from 0 to 200 in increments of 10. Search tasks are generated according to predefined data ratios.
		\begin{figure}[!t]
		\centering
		\includegraphics[width=2.5in]{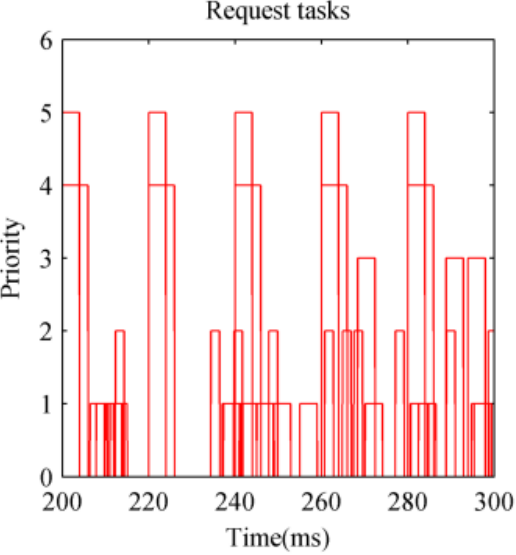}
		\caption{Partial request sequence diagram of the original tasks.}
		\label{fig:6}
		\end{figure}

The radar detection range is assumed to be 50-300 km, with a simulation time of 0-12 seconds. The phased array radar scheduling interval is set to SI = 50 ms, and targets appear randomly. Due to the random generation of targets, the tracking task exhibits a degree of randomness. The number of precision tracking targets increases by 10 with each iteration, and the simulation is repeated 100 times. The parameters for the radar task types are summarized in Table I.

The simulation results are compared with those from previous works, including traditional time-hand-based methods \cite{[9]Lu2006}, \cite{[12]Zhao2012}, and \cite{[21]Zhang2017}. A section of the original task request sequence, from 200 ms to 300 ms, is extracted for analysis, as shown in Figure \ref{fig:6}.

\begin{table*}[!t]
\centering
\caption{Radar Task Type Parameter Table}
\label{tab:radar_task_type}
\begin{tabular}{|c|c|c|c|c|c|}
    \hline
    Number & Radar task type & Priority & Beam resident parameter $(t_x, t_w, t_r)$/ms & Time window & Replacement ratio \\
    \hline
    1 & Low priority search & 1 & 1.0, 2.0, 1.0 & 100ms & 50Hz \\
    2 & High priority search & 2 & 0.5, 1.5, 0.5 & 50ms & 50Hz \\
    3 & General tracking & 3 & 0.5, 0.9, 0.5 & 50ms & 2Hz \\
    4 & Precision tracking & 4 & 0.5, 0.5, 0.5 & 30ms & 5Hz \\
    5 & Verify & 5 & 1.0, 1.5, 1.0 & 30ms & 2Hz \\
    \hline
\end{tabular}
\end{table*}

\subsection{Simulation Results.}
		\begin{figure}[!t]
		\centering
		\includegraphics[width=3in]{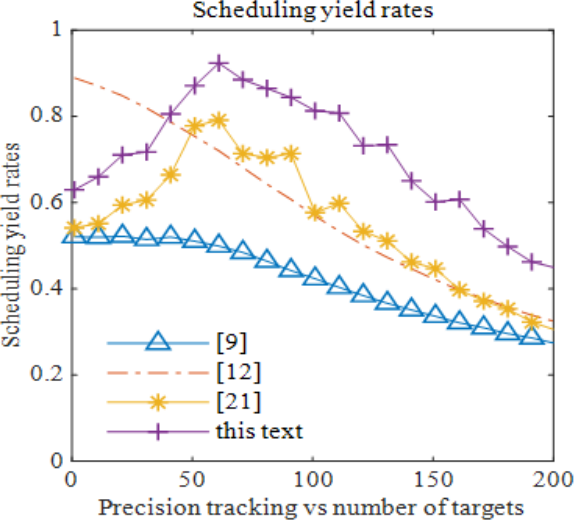}
		\caption{Scheduling Yield Ratios.}
		\label{fig:7}
		\end{figure}

		\begin{figure}[!t]
		\centering
		\includegraphics[width=3in]{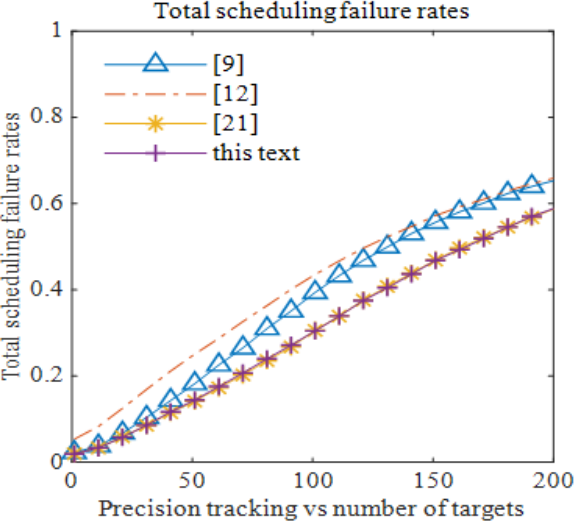}
		\caption{Total Scheduling Failure Ratios.}
		\label{fig:8}
		\end{figure}

As shown in Fig. 6 and Fig. 7, the scheduling yield of the dynamic pulse interleaving algorithm is not only higher than that of the traditional time-pointer algorithm presented in \cite{[9]Lu2006}, but also exceeds the yield of the quadratic programming algorithm in \cite{[12]Zhao2012}. Additionally, the dynamic pulse interleaving algorithm achieves a higher scheduling yield compared to the pulse interleaving algorithm in Zhang’s work \cite{[21]Zhang2017}.

When the number of precision tracking targets is below 50, the original quadratic programming algorithm achieves the highest yield, outperforming other algorithms, with the pulse interleaving algorithm ranking second. In this scenario, the scheduling gain is greater, and the scheduling failure ratio is the lowest. As the number of precision tracking targets increases, the dynamic pulse interleaving algorithm shows the highest scheduling yield ratio, surpassing both the traditional pulse interleaving algorithm and other methods. Furthermore, the overall scheduling failure ratio of the dynamic pulse interleaving algorithm remains consistently lower than that of other algorithms.
		\begin{figure}[!t]
		\centering
		\includegraphics[width=3in]{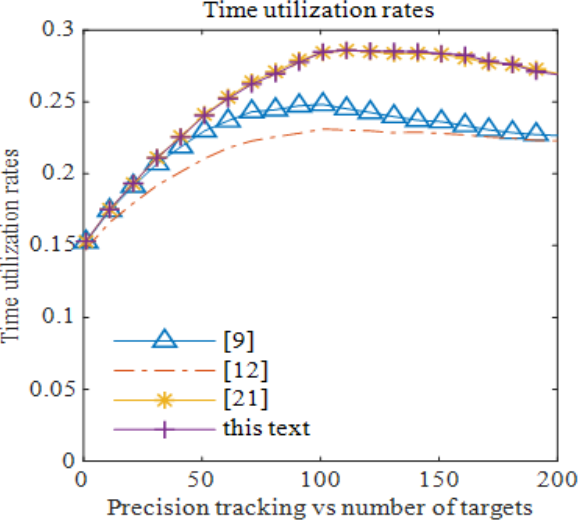}
		\caption{Time Utilization.}
		\label{fig:9}
		\end{figure}

		\begin{figure}[!t]
		\centering
		\includegraphics[width=3in]{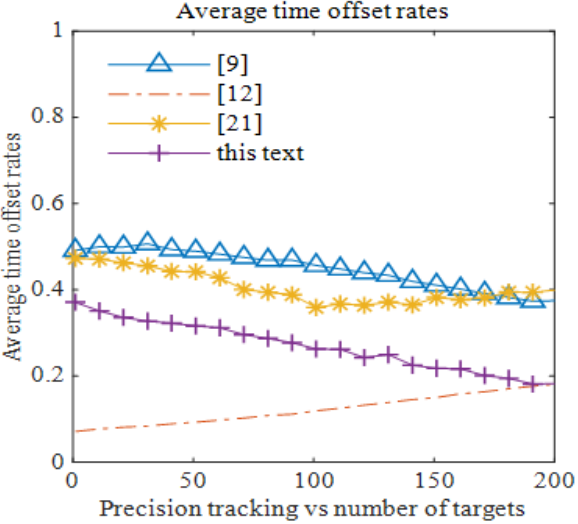}
		\caption{Variation of Average Time shift Ratios.}
		\label{fig:10}
		\end{figure}

As shown in Fig. 8, the time utilization ratio of the pulse interleaving algorithm proposed in this paper is consistent with that of the method in \cite{[21]Zhang2017}, and it surpasses other algorithms in terms of time utilization. In Fig. 9, it can be observed that the time shift ratio of the proposed method, which incorporates task synthesis priority with time shift considerations, is lower than other algorithms, and only slightly higher than that of the quadratic programming algorithm in \cite{[12]Zhao2012}. As the number of precision tracking targets increases, the time shift ratio of the proposed method approaches that of \cite{[12]Zhao2012}, further enhancing the effectiveness of task execution.

The pulse interleaving algorithm presented in this paper shows similar performance to \cite{[21]Zhang2017} in terms of scheduling time utilization and overall failure ratios. However, the time shift ratio is reduced by 35\%, and the scheduling gain is improved by 30\%. Compared to \cite{[9]Lu2006}, the time utilization ratio is improved by nearly 20\%, the time shift ratio is reduced by 40\%, and the scheduling benefit is increased by 45\%.
		\begin{figure}[!t]
		\centering
		\includegraphics[width=3in]{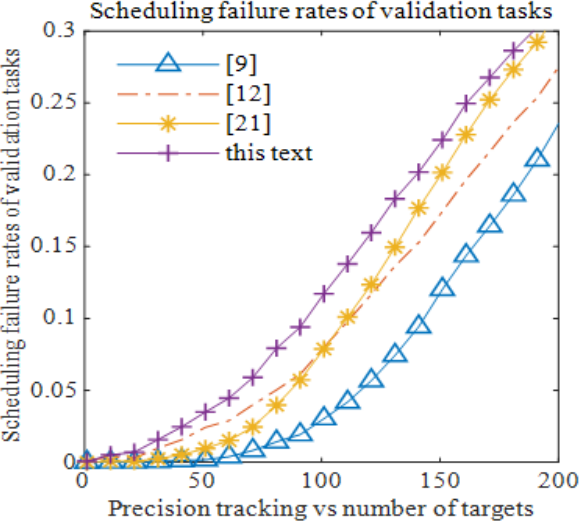}
		\caption{Scheduling Failure Ratios of Validating Tasks.}
		\label{fig:11}
		\end{figure}

		\begin{figure}[!t]
		\centering
		\includegraphics[width=3in]{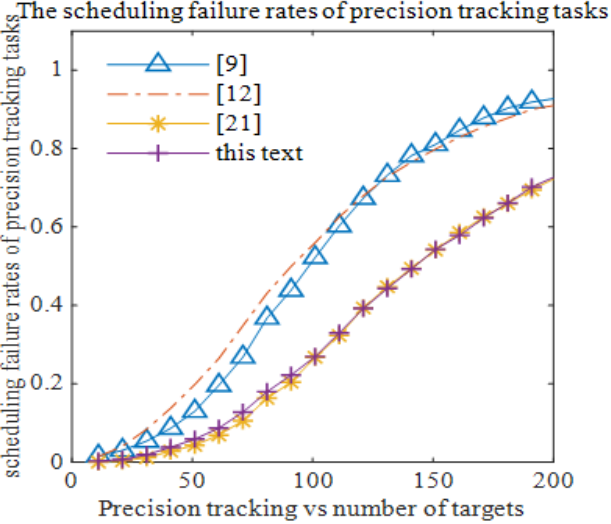}
		\caption{Scheduling Failure Ratios of Precision Tracking tasks.}
		\label{fig:12}
		\end{figure}

		\begin{figure}[!t]
		\centering
		\includegraphics[width=3in]{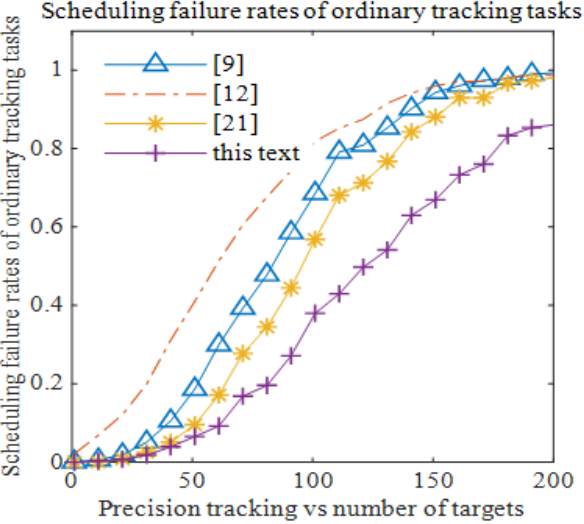}
		\caption{Scheduling Failure Ratios of Ordinary tracking tasks.}
		\label{fig:13}
		\end{figure}

		\begin{figure}[!t]
		\centering
		\includegraphics[width=3in]{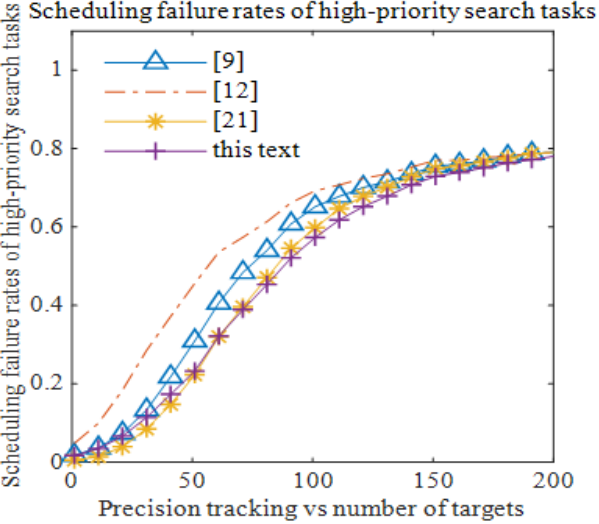}
		\caption{Scheduling Failure Ratios of High Priority Search Tasks.}
		\label{fig:14}
		\end{figure}

		\begin{figure}[!t]
		\centering
		\includegraphics[width=3in]{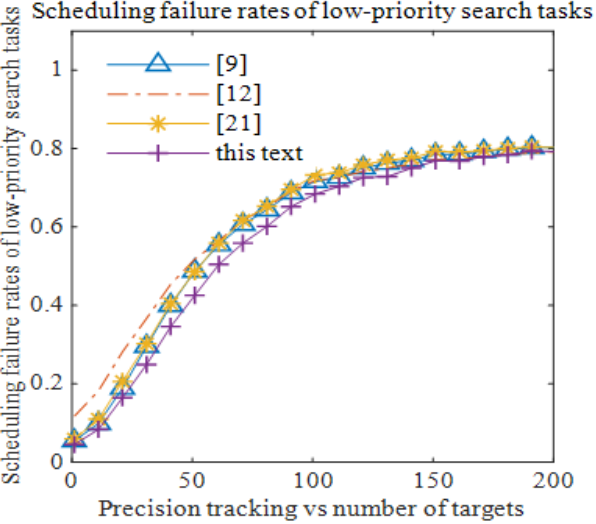}
		\caption{Scheduling Failure Ratios of Low Priority Search Tasks.}
		\label{fig:15}
		\end{figure}

As shown in Fig. 10, the scheduling failure ratios for verification tasks using this method are similar to those of other scheduling algorithms. However, Figs. 11, 12, 13, and 14 demonstrate that the proposed algorithm achieves the lowest scheduling failure ratios across precision tracking, general tracking, high-priority search, and low-priority search tasks. Notably, the failure ratio for general tracking task scheduling is significantly lower than that of other algorithms.

\section{Conclusion}
\noindent This paper addresses the resource scheduling problem of phased array radar, with a particular focus on classifying tracking task modes based on target threat levels. The proposed method employs a dynamic priority assignment strategy and a pulse interleaving scheduling algorithm, which takes into account task working modes, deadlines, and time shift ratios. This approach not only allows for the implementation of different scheduling strategies by emphasizing various task characteristics but also ensures the uniqueness of task priorities. As a result, more tasks can be executed when high-priority search tasks are scheduled. The use of pulse interleaving enhances both the flexibility of task scheduling and the utilization of time resources.

In summary, the proposed synthesis priority and pulse interleaving algorithm effectively reduces the time shift ratio and improves task scheduling efficiency compared to Zhang's algorithm. The proposed algorithm achieves lower task scheduling failure ratios, reduced time shift ratios, higher scheduling yields, and improved time utilization ratios. Specifically, compared to Zhang's algorithm, the time shift ratio is reduced by 35\%, and the scheduling yield is increased by 30\%. Compared to the traditional time-pointer algorithm, the time utilization ratio is nearly 20\% higher, the overall task scheduling failure ratio is reduced by 10\%, the time shift ratio is 40\% lower, and the scheduling yield is 45\% higher. Furthermore, compared to the quadratic programming algorithm, the proposed method also exhibits higher scheduling yields, better time utilization, lower task scheduling failure ratios, and reduced time shift ratios.

Simulation results demonstrate that the proposed algorithm outperforms other algorithms in terms of scheduling performance.

\end{document}